\documentclass[runningheads]{llncs}
\usepackage[T1]{fontenc}
\usepackage{graphicx}
\usepackage{tikz}

\usepackage{color}
\usepackage{amssymb}
\usepackage{amsmath}
\usepackage[markup=underlined]{changes}
\usepackage{comment}
\usepackage{todonotes}
\setcommentmarkup{\todo[color={authorcolor!20},size=\scriptsize]{#3: #1}}

\def\eps{\varepsilon}
\newcommand{\ra}{\rightarrow}

\newcommand{\bean}{\begin{eqnarray*}}
	\newcommand{\eean}{\end{eqnarray*}}

\newcommand{\bp}{\begin{proof}}
	\newcommand{\ep}{\qed\end{proof}}
\def\bbbn{{\rm I\!N}}
\newcommand{\N}{{\bbbn}}
\def\bigo{\mathcal{O}}

\begin{document}
\title{Jump Complexity of Deterministic Finite Automata with Translucent Letters\thanks{This study was supported by PNRR/2022/C9/MCID/I8 project 760096. It was also performed through the Core Program within the National Research, Development and Innovation Plan 2022-2027, carried out with the support of MRID, project no. 23020101(SIA-PRO), contract no 7N/2022, and project no. 23020301(SAFE-MAPS), contract no 7N/2022. The first author was supported by JSPS KAKENHI Grant Number 23K10976.}}
\titlerunning{Jump Complexity of DFAwtl}
%

\author{Szil\'ard Zsolt Fazekas \and
Victor Mitrana\inst{2} \and
Andrei P\u{a}un\inst{3,4} \and
Mihaela P\u{a}un\inst{3,4}}
\authorrunning{S.Z. Fazekas et al.}
%
\institute{Akita University, Japan\\
\email{szilard.fazekas@ie.akita-u.ac.jp} \and
Universidad Politecnica de Madrid, Spain\\
\email{victor.mitrana@upm.es} \and
National Institute of Research and Development for Biological Sciences, Romania
\email{andrei.paun@incdsb.ro} \and
University of Bucharest, Romania\\
\email{mihaela.paun@incdsb.ro}}
\maketitle              

\begin{abstract}
We investigate a dynamical complexity measure defined for finite automata with translucent letters (FAwtl). Roughly, this measure counts the minimal number of necessary jumps for such an automaton in order to accept an input.
The model considered here is the deterministic finite automaton with translucent letters (DFAwtl). Unlike in the case of the nondeterministic variant, the function describing the jump complexity of any DFAwtl is either bounded by a constant or it is linear. We give a polynomial-time algorithm for deciding whether the jump complexity of a DFAwtl is constant-bounded or linear and we prove that the equivalence problem for  DFAwtl of $\bigo(1)$ jump complexity is decidable. We also consider another fundamental problem for extensions of finite automata models,
deciding whether the language accepted by a FAwtl is regular. We give a positive partial answer for DFAwtl over the binary alphabet, in contrast with the case of NFAwtl, where the problem is undecidable.
\end{abstract}

\section{Introduction}
The finite automaton is a cornerstone of theoretical computer science, providing a fundamental model for understanding regular languages. Over the years, various extensions of finite automata have been proposed to increase their expressive power and applicability in various areas. Some of those extensions attempted to introduce a non-sequential processing of the inputs, where read letters are consumed from the tape, and the automaton is allowed to jump over certain parts of the input or shift the remaining input word. These latter types of models include the input-revolving automaton~\cite{inputrev}, the jumping automaton~\cite{jumping} or the one-way~\cite{onewayjumping} and two-way jumping automaton~\cite{two-way}.

The Finite Automaton with Translucent Letters (FAwtl)~\cite{translucent} is such an extension of classical finite automata (FA). The FAwtl introduces the concept of translucency, allowing certain input symbols to be skipped over in particular states. Translucent letters in FAwtl enable the automaton to "ignore" certain symbols in specified states, effectively jumping to the next non-translucent symbol. This feature extends the classical finite automata model by introducing a controlled form of non-sequential input reading and it enhances the computational power of FAwtl, enabling them to recognize patterns that are beyond the reach of traditional FA, while inheriting many good algorithmic properties from them. A recent survey by Otto~\cite{survey} provides a detailed picture of variants of the model and their computational power and closure properties. 

The jump complexity of a FAwtl refers to the number of jumps required during the processing of an input string in the worst case, for each length. Understanding this complexity seems crucial for evaluating the computational efficiency and language recognition power of FAwtl, as well as related models, like the one-way jumping FA~\cite{onewayjumping}. This paper explores the jump complexity of deterministic FAwtl and its implications for the decidability of language equivalence.

In the broader context of automata theory, the study of jump complexity aligns with ongoing research into non-regular complexity measures for models such as extended automata over groups~\cite{jlamp} (group memory operations), one-way jumping automata~\cite{jump,sweep} (jump and sweep complexity), and even context-free grammars (number of applications of non-regular rules), or push-down automata~\cite{degree} (number of push operations). The study of the complexity classes defined through those models sharpens the picture of how complicated languages are at the lower end of the computational complexity hierarchies and, in particular, seeks to understand the threshold in the amount of non-regular resources to cross the boundary between regular and non-regular language classes.

The notion of jump complexity for nondeterministic FAwtl (NFAwtl) has been recently investigated~\cite{MitranaPPS24}, establishing a detailed hierarchy of separable complexity classes $\bigo(1)$, $\bigo(\log n)$, $\bigo(\sqrt{n})$ and $\bigo(n)$. It was shown that NFAwtl with $\bigo(1)$ jump complexity accept regular languages, directly implying the same conclusion for deterministic FAwtl (DFAwtl). This provides a sufficient condition for regularity, though it is not a necessary one. Some FAwtl with $\omega(1)$ jump complexity can still recognize regular languages, also opening the possibility to study descriptional complexity tradeoffs versus classical DFA/NFA.

The objective of this paper is to provide a detailed analysis of the jump complexity of DFAwtl. We settle the topic by showing that the complexity hierarchy collapses to the two extremes $\Theta(1)$ and $\Theta(n)$, and we give a polynomial time algorithm that decides for a given DFAwtl which extreme its jump complexity falls into. We also demonstrate the applications of our approach to achieve positive results regarding two of the most fundamental problems in the study of extensions of finite automata: decidability of language equivalence and decidability of regularity.

\section{Definitions and Notations}
The basic concepts and notations that are to be used throughout the paper are defined in the sequel; the reader may consult \cite{handbook} for basic concepts that are not defined here. 

We will denote by $\mathbb{N}$ the set of natural numbers starting from $1$ and $\mathbb{N}_0=\mathbb{N}\cup \{0\}$. For all $i,j\in \mathbb{N}_0$ with $i\leq j$, we denote the closed interval between $i$ and $j$ by $[i,j]=\{i,\dots,j\}$. For succinctness we also use $[n]=[0,n]$.

In addition, we use the following concepts and notations:
\begin{itemize}
    \item $\#(X)$ is the cardinality (number of elements) of the finite set $X$;
    \item $V^*$ is the set of all finite words formed by symbols in the alphabet $V$;
    \item $|x|$ is the length of word $x$;
    \item $|x|_U$ is the length of the word obtained from $x$ by erasing all letters that are not in $U$; when $U$ is a singleton $\{a\}$, we sometimes write $|x|_a$;
    \item $\eps\in V^*$ is the empty word, $|\eps|=0$;
    \item for $x\in V^*$, $alph(x)$ is the minimal alphabet $U\subseteq V$ such that $x\in U^*$.
\end{itemize}

A \textit{nondeterministic finite automaton} (NFA) 
is a quintuple $M=(Q,V,\delta$, $q_0,F)$, where $Q$ is a finite set of states, $V$ is an alphabet disjoint from $Q$, $\delta$ is a mapping from $Q \times V$ into $2^Q$, called the transition mapping,
$q_0 \in Q$ is the initial state, and $F\subseteq Q$ is the set of final states. 

A configuration of $M$ is a pair in $Q \times V^{*}$. A transition relation is defined on the set of configurations of $M$ as follows:
$(s,ay) \rightarrow (p,y)$ if $p\in \delta(s,a)$, $s,p\in Q$, $a\in V$, $y\in V^*$.
The reflexive and transitive closure of the relation $\rightarrow$ is denoted 
by $\rightarrow^*$. The language accepted by $M$ is
$L(M)=\{w\in V^*\mid (q_0,w)\rightarrow^* (s,\eps), s\in F\}$.

$M$ is a \textit{deterministic finite automaton} (DFA) if for each $s\in Q$ and each $a \in V$, there is at most one state in $\delta(s,a)$. 

A {\it nondeterministic finite automaton with translucent letters} (NFAwtl) is an NFA
$M$ as above, such that the transition relation is defined in the following way.
First, we define the partial relation $\circlearrowright$ on the set of all
configurations of $M$: $(s,xay) \circlearrowright (p,xy)$ iff $p\in\delta(s,a)$,
and $\delta(s,b)$ is not defined for any $b\in alph(x)$, $s,p\in Q$,
$a,b\in V$, $x\in V^+$, $y\in V^*$.
We now write 
$$(p, x)\models_M (q, y), \mbox{ if either } (p, x)\ra (q, y) \mbox{ or } (p, x) \circlearrowright (q, y).$$  
The subscript $M$ is omitted when it is understood from the context.
 
The language accepted by $M$ is defined by 
$$L(M)=\{x\in V^*\mid (q_0, x)\models^* (f,\eps), f\in F\}.$$
It is worth mentioning that the NFAwtl was introduced in \cite{translucent}, with a slightly different definition. Our definition is a variant of  an NFAwtl in the normal form in \cite{translucent} without a marker for the end of the input word, following the definition used in the study of jump complexity of NFAwtl~\cite{MitranaPPS24}. Actually, a more in-depth discussion is necessary here. The original model allows multiple initial states. This feature is dropped in our version, and it is not particularly relevant for us, as our results only tangentially mention the nondeterministic machines. The translucent function is missing but this can be replaced as follows: (i) no transition is defined for a state and its translucent letters, and (ii) a deadlock state is defined for any state $q$ and any symbol $a$ that is not in the set of translucent letters of $q$ and there is no transition for the pair $(q,a)$ in the original automaton.
Furthermore, the original automaton could accept the input without reading all of it, but this is not possible in our model. In spite of all these differences, we preferred to keep the original name of the model. An alternative name could be {\it one-way returning jumping finite automaton}, as it resembles features from the one-way jumping automata model as well. As the model considered here is not exactly the original NFAwtl, we will make use of results known for the original NFAwtl only if the result remains valid for
our model.

This automaton is also related to the {\it one way jumping automaton} introduced in \cite{onewayjumping} with the difference that after each jump it returns to its previous position and does not shift
the jumped part to the end of the word. 

The condition for a finite automaton with translucent letters to be deterministic (DFAwtl) is exactly the same to that for a finite automaton.

To simplify referencing states that jump over certain letters, we introduce the following terminology. For a DFAwtl $A=(Q,\Sigma,q_0,\delta,F)$ and a non-empty set of letters $\Gamma\subset \Sigma$, we say that a state $q\in Q$ is $\Gamma$-deficient if for each letter $a\in \Gamma$, the transition $\delta(q,a)$ is undefined. A state is \emph{deficient} if it is $\Gamma$-deficient for some non-empty $\Gamma\subset \Sigma$. 

We continue by introducing the complexity measure that is the main focus of this paper.
Let $M$ be an NFAwtl; we consider $w\in L(M)$, and the accepting computation in $M$ on the input $w$:
$$C_M(w): (q_0, w)\models (q_1, w_1)\models (q_2, w_2)\models\dots \models (q_m,\eps),$$ with $q_i\in Q$, $1\le i\le m$, and $q_m\in F$. 
We define 
$$\varphi(C_M(w))=\{i\mid (i\ge 1) \& ((q_{i-1}, w_{i-1})  \circlearrowright (q_i, w_i))\}.$$
In words, $\varphi(C_M(w))$ contains all the jumping steps in the computation $C_M(w)$.

We now define the 
{\it jump complexity} of the computation of $M$ on the word $w$ by
$$jc_M(w)= \left\{
\begin{array}{ll}
\min \{\#(\varphi(C_M(w)))\mid C_M(w) \mbox{ is an accepting computation}\}\\
\mbox{\underline{undefined}, if } w\notin L(M).
\end{array}
\right.$$

In other words, the jump complexity of an accepted word with respect to $M$ is computed by taking into consideration the “least non-regular computation”, if there is one. Equivalently, the jump complexity of a word $w\in L(M)$ with respect to $M$ is the number of jumping steps of an accepting computation of $M$ on $w$ with the minimal number of jumping steps. If the word is not accepted by the automaton, then its jump complexity is undefined. 

The jump complexity of an automaton $M$ as above is a partial mapping from $\N$ to $\N$ defined by
$$JC_M(n) =\max\{jc_M(w) | |w|=n, w \in L(M)\}.$$
As one can see, the most ``non-regular'' word of each length in the accepted language is considered. If $L(M)$ does not contain any word of some length $n$, then $JC_M(n)$ is undefined.
It is clear that $JC_M(n)\le n$, for every NFAwtl $M$, as one letter is consumed in every step of a computation.

Let $f$ be a function from $\N$ to $\N$; we define the family of languages
\bean 
JCL(f(n))&=&\{L\mid \exists \mbox{ NFAwtl $M$ such that } L=L(M) \mbox{ and }\\ && JC_M(n)\in \mathcal{O}(f(n))\}.
\eean
The above definitions naturally apply to DFAwtl, where the least non-regular computation for any input is the only computation performed by the machine on that input.

We start with an example which will be used later.
Consider the non-regular language $L=\{w\in\{a,b\}^*\mid |w|_a=|w|_b=n\ge 0\}$.
The very simple DFAwtl $M$, depicted in Figure 1, accepts this language.

\begin{figure}[h]	
	\centering
	\includegraphics[width=6cm]{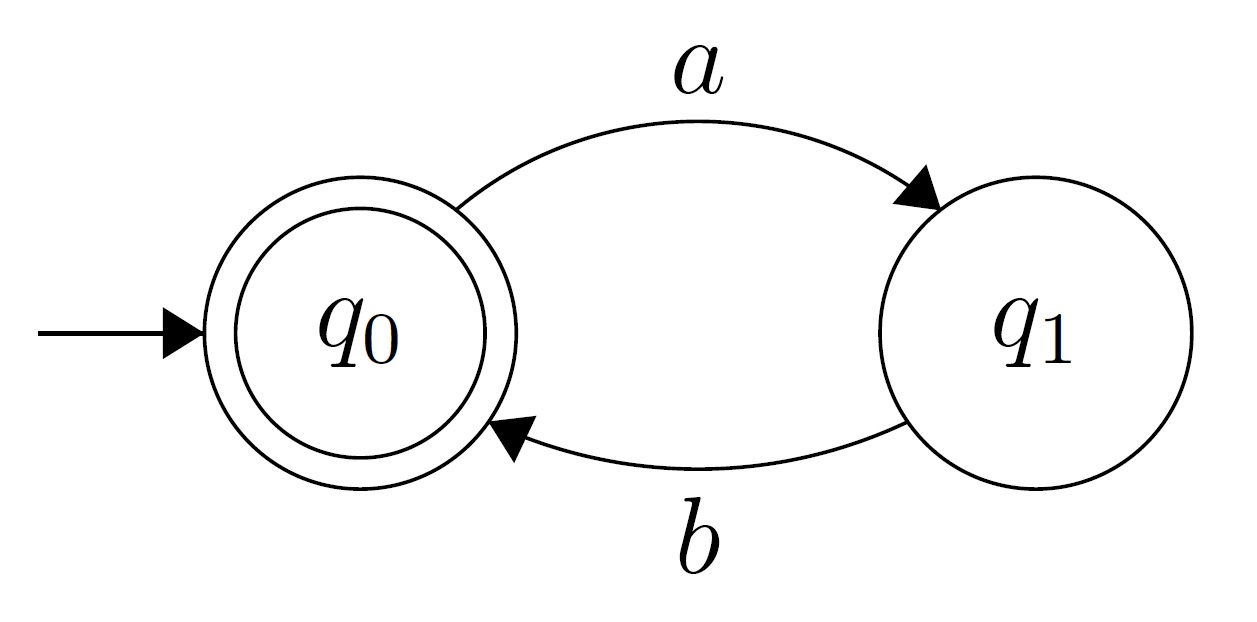}
        \caption{A DFAwtl accepting $L$.}
	\label{Fig1}
\end{figure}

Indeed, the automaton accepts the language $L=\{w\in\{a,b\}^*\mid |w|_a=|w|_b=n\ge 0\}$ by the following simple observation. Whenever the automaton is in the initial state, it has to consume an $a$ (no matter if that $a$ is the current symbol or the automaton must do a jump) in order to reach $q_1$. As $q_1$ is not a final state, the automaton must consume 
a $b$ for returning to the initial state, which is also final. Therefore, the number of occurrences of $a$'s and $b$'s must be the same.

On the other hand, it is easy to check that $0\le jc_M(w)\le n$ for any $w\in L$ of length $2n$. 
As explained above, for going out from and returning to $q_0$, the automaton consumes an $a$ and a $b$, with at most one jump. For $jc_M(b^na^n)=n$, it follows that $JC_M(n)$ is either $n/2$, if $n$ is even, or
undefined, otherwise. From here, $L\in JCL(n/2)$ holds.

As the algorithms presented later reference the graph, or digraph of the automata considered, we briefly define that notion. An (unlabeled) digraph is a pair $G=(V,E)$ where $V$ is a finite set of \emph{vertices} and $E\subseteq V\times V$ is a set of ordered pairs of vertices, called \emph{edges} or \emph{arcs}.

\section{Results}

\begin{theorem}\label{decidable1}
	Given a DFAwtl $M$ it is algorithmically decidable whether or not the jump complexity of $M$ is bounded by a constant. 
\end{theorem}

\bp
Let $M=(Q,V,\delta,q_0,F)$ be a DFAwtl. Without loss of generality we may assume that 
all the states of $M$ are useful, namely for each state $q\in Q$ there exist $x,y\in V^*$ such that $(q_0,x)\rightarrow^* (q,\eps)$ and $(q,y)\rightarrow^* (f,\eps)$, with $f\in F$. It is an easy exercise to remove all the useless states exactly as it is done for a finite automaton.

We start with an important remark. Let us consider the following computation in $M$:
$$(q_0,w_1w_2)\rightarrow^* (q,w_2)\circlearrowright (s,w_2')\models^* (f,\eps),$$
with $f\in F$, and $jc_M(w_1w_2)=n$, for some $n$.
We deduce that the next computation is also valid in $M$:
$$(q_0,w_1 a w'_2)\rightarrow^* (q,aw_{21}w_{22})\rightarrow (s,w_{21}w_{22})\models^* (f,\eps),$$
where $\delta(q,b)=\emptyset$ for all $b\in alph(w_{21})$, $\delta(q,a)=s$,
$w_2'=w_{21}w_{22}$, and $w_2=w_{21}aw_{22}$. Informally, we moved the letter consumed by $M$ when jumping from $q$ in the former computation as the current symbol in the latter computation. Clearly, $jc_M(w_1aw_2')=n-1$ holds.
Actually, this fact may be extended such that starting from a given a partial valid computation with $n$ jumps, one can construct a partial valid computation with $n-1$ jumps.

For each pair $q\in Q, a\in V$ such that $\delta(q,a)=\emptyset$, we construct
the set
\bean
\langle q,a\rangle &=& \{s\in Q\mid \mbox{ there exist } x\in V^+, y\in V^* \mbox{ such that } (q,x)\rightarrow^* (s,\eps) \mbox{ and } \\
&& (s,ax)\models^* (q,\eps)\}.
\eean


\noindent {\it \bf Claim 1.} If there exists a pair $(q,a)$ such that $\delta(q,a)=\emptyset$ and $\langle q,a\rangle \ne\emptyset$, then $JC_M(n)$ is not a constant-bounded function.\\
{\it \bf Proof of Claim 1.} 
Let $\delta(q,a)=\emptyset$, $\delta(q,b)=s$, and 
$(s,ax)\rightarrow^* (q,\eps)$. Furthermore, let $y,z\in V^*$ such that
$(q_0,y)\rightarrow^* (q,\eps)$ and $(q,z)\rightarrow^* (f,\eps)$, with $f\in F$.
We now consider the words
$$w_n=y(abx)^nz,$$
for all $n\ge 1$. Obviously, all these words are accepted by $M$, and $jc_M(w_n)\ge n$,
for all $n\ge 1$. As the automaton is deterministic, these computations are unique
for each $w_n$, which concludes the proof of Claim 1.

\noindent {\it \bf Claim 2.} If
we have a state $q$ and letter $a$ such that the following conditions hold, then $JC(M)$ is not constant-bounded:
\begin{itemize}
    \item $\delta(q,a)=\emptyset$;
    \item there is some word $x$ such that $(q,x)\rightarrow^* (q,\varepsilon)$;
    \item for each prefix $x'$ of $x$, the state $\delta(q,x')$ is $a$-deficient;
    \item there are some words $y$ and $z$ (with $y$ not containing $a$ and $z$ possibly
empty) such that $(q,yaz)\rightarrow^* (f,\varepsilon)$, for some final state $f$, and for each prefix $y'$ of $y$, the state $\delta(q,y')$ is $a$-deficient.

\end{itemize}
{\it \bf Proof of Claim 2.}
If those conditions hold, then words of the form $u a x^n yz$ are
accepted with at least $n$ jumps, so the jump complexity is linear (where $u$
is a word such that $(q_0,u)\rightarrow^* (q,\varepsilon)$).

\noindent {\it \bf Claim 3.} If all the sets $\langle q,a\rangle$ computed above are empty, and the conditions in Claim 2 are not satisfied, then $JC_M(n)$ is a constant-bounded function. \\
{\it \bf Proof of Claim 3.} 
Assume by contradiction that for any constant $c$, there exists a word $w_c$ in $L(M)$ such that $jc_M(w_c)>c$, and take $c$ to be the cardinality of $Q$. By the pigeonhole principle, this means that at least two of the jumps happen in the same state $q$ and that in the transition graph of $M$ there is a cycle starting and ending in $q$.
More formally, if the jump complexity of $M$ is not bounded by any constant, then there must be deficient states $q$ for which there exists some word $x$
with $(q,x)\rightarrow^* (q,\varepsilon)$, and any computation with more than $|Q|$ jumps must go through such a state $q$ at least twice. For each such $q\in Q$, $a\in V$ and $x\in V^*$ such that $\delta(q,a)=\emptyset$ and $(q,x)\rightarrow^* (q,\varepsilon)$:
\begin{itemize}
    \item if $x$ contains $a$, then $\langle q,a\rangle\neq \emptyset$, so the condition of Claim 1 is met;
    \item if all prefixes $x'$ of $x$ are such that $\delta(q,x')$ is $a$-deficient, then the conditions of Claim 2 are met.
\end{itemize}

Now let us assume that the conditions of Claim 1 and Claim 2 do not hold. Furthermore, let $q$ be a state in which at least two jumps are performed during the computation with more than $|Q|$ jumps. When the first jump in state $q$ happens, let us assume that the letter jumped over was $a$, that is, $a$ was the first letter of the remaining input and $q$ is $a$-deficient. Depending on when the $a$ that was jumped over was read, we distinguish the following cases:
\begin{enumerate}
    \item the $a$ was read between the two jumps from $q$: this means that there is some cycle in $q$ labeled by a word containing $a$, satisfying the conditions of Claim 1, a contradiction.
    \item the $a$ was read after the second jump from $q$: this means all states visited between the two jumps from $q$ are $a$-deficient, but that implies the existence of a cycle in $q$ with only $a$-deficient states in it, satisfying the conditions of Claim 2, again a contradiction.
    
\end{enumerate}

From the contradictions above we can deduce that the conditions of Claim 1 or Claim 2 are necessary for $M$ to have jump complexity not bounded by any constant, and earlier we have seen that they are sufficient, too, concluding the proof.

By these claims it follows that the algorithm for deciding the stated problem
consists in computing all the sets $\langle q,a\rangle$, $q\in Q$, $a\in V$, and checking the existence of reachable cycles that consist of $a$-deficient states for some $a$, and from which there is a path to a final state containing at least one $a$ transition. It is straightforward that the checks above are effective, e.g., by a brute force search of the transition graph of $M$, but as we will see later, one can do them efficiently.
\ep

It is worth mentioning that the conditions for $M$ to have jump complexity not included in $\bigo(1)$ can be easily checked by using algorithms for searching paths in a directed graph, see, e.g., \cite{Floyd}. More precisely,

\begin{theorem}\label{complexity} 
The complexity of the algorithm in the proof of Theorem \ref{decidable1} is \\
$\mathcal{O}(\#(Q^3)\cdot \#(V))$, where $Q$ and $V$ is the set of states and the alphabet of $M$, respectively.
\end{theorem}

\bp
For each pair $q\in Q, a\in V$ such that $\delta(q,a)=\emptyset$, besides the definition of the set $\langle q,a\rangle$, we construct the set
\bean
[ q,a ] &=& \{s\in Q\mid \mbox{ there exist } x\in V^+, y\in V^* \mbox{ such that } (q,x)\rightarrow^* (s,\eps) \mbox{ and } \\
&& (s,ax)\rightarrow^* (q,\eps)\}.
\eean
Clearly, $[q,a]\subseteq \langle q,a\rangle$, for all pairs $(q,a)$ such that $\delta(q,a)=\empty$. For simplicity, we set $[q,a]=\langle q,a\rangle=\emptyset$, for all pairs $(q,a)$ such that $\delta (q,a)\ne\emptyset$.

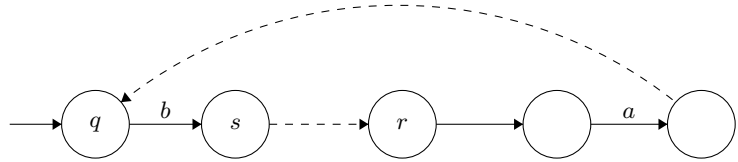
\begin{figure}
    \begin{center}
\begin{tikzpicture}[scale=0.15]
\tikzstyle{every node}+=[inner sep=0pt]
\draw [black] (12,-24.5) circle (3);
\draw (12,-24.5) node {$q$};
\draw [black] (24.4,-24.5) circle (3);
\draw (24.4,-24.5) node {$s$};
\draw [black] (39.2,-24.5) circle (3);
\draw (39.2,-24.5) node {$r$};
\draw [black] (52.9,-24.5) circle (3);
\draw [black] (65.7,-24.5) circle (3);
\draw [black] (4.4,-24.5) -- (9,-24.5);
\fill [black] (9,-24.5) -- (8.2,-24) -- (8.2,-25);
\draw [black] (15,-24.5) -- (21.4,-24.5);
\fill [black] (21.4,-24.5) -- (20.6,-24) -- (20.6,-25);
\draw (18.2,-24) node [above] {$b$};
\draw [black] (55.9,-24.5) -- (62.7,-24.5);
\fill [black] (62.7,-24.5) -- (61.9,-24) -- (61.9,-25);
\draw (59.3,-24) node [above] {$a$};
\draw [black,dashed] (27.4,-24.5) -- (36.2,-24.5);
\fill [black] (36.2,-24.5) -- (35.4,-24) -- (35.4,-25);
\draw [black] (42.2,-24.5) -- (49.9,-24.5);
\fill [black] (49.9,-24.5) -- (49.1,-24) -- (49.1,-25);
\draw [black,dashed] (14.273,-22.543) arc (128.55255:51.44745:39.435);
\fill [black] (14.27,-22.54) -- (15.21,-22.44) -- (14.59,-21.65);
\end{tikzpicture}
\end{center}
    \caption{If $\langle q,a\rangle$ is not empty, then there exists $r$ such that $[r,a]$ is not empty.}
    \label{fig:shift}
\end{figure}

\noindent {\it \bf Claim 1.} All the sets $[q,a]$ are empty if and only if all the sets $\langle q,a\rangle$ are empty.\\
{\it \bf Proof of Claim 1.} It suffices to prove that all the sets $\langle q,a\rangle$ are empty by assuming that all the sets $[q,a]$ are empty.
Let us assume that $\delta(q,a)=\emptyset$ and the set $\langle q,a\rangle$ is not empty for some state $q$ and letter $a$. Let $s\in \langle q,a\rangle$, which means that $\delta(q,b)=s$ and $(s,ax)\models^* (q,\eps)$, for some letter $b$ and string $x$. We argue about the states in this cycle beginning and ending in $q$. 
If $\delta(s,a)\neq \emptyset$, then $s\in [q,a]$, which contradicts $[s,a]$ being empty. If $\delta(s,a)=\emptyset$, then $\langle s,a\rangle\neq \emptyset$ and we try to show that $[s,a]\neq \emptyset$, shifting the argument to the next state in the cycle. If we keep shifting the state in the cycle, eventually we reach the state $r$ from which the last jump is performed before reading $a$ sequentially, that is, $\langle r,a\rangle\neq \emptyset$ and $[r,a]\neq \emptyset$. Such a state $r$ must exist, because the $a$ is read in the cycle, by the definition of $\langle q,a\rangle$, contradicting the assumption that $[r,a]$ is empty (see Fig.~\ref{fig:shift}).
\color{black}

We now check the conditions of the claims 1 and 2 from the previous proof.
For checking the condition of Claim 1 in that proof, we actually check the equivalent condition proved in Claim 1 above.

 Formally, we consider the unlabeled digraph of the automaton $M$ and construct the transitive closure of this graph. As it is known, see, for instance, the Floyd-Warshall algorithm, this can be done in $\mathcal{O}(\#(Q^3))$.
Now, for each pair $(q,a)\in Q\times V$, we construct the following sets:
\begin{itemize}
    \item $X=\{q_1,q_2,\dots,q_p\}\subseteq Q$, where each $q_i$, $1\le i\le p$, is reachable from $q$ in $M$ in one step.
This step requires $\mathcal{O}(\#(Q))$ time.
    \item $Y=\{r_1,r_2,\dots,r_m\}\subseteq Q$, where each element of $Y$ is reached from some element of $X$ by reading $a$. If $q\in Y$, then $[q,a]\ne\emptyset$,
and the algorithm halts. This step can be accomplished in $\mathcal{O}(\#(Q^2))$ time.
\end{itemize}

By using the transitive closure computed above, we deduce that
$[s,a]$ is empty if and only if $s$ is not reachable from any state in $Y$. By these explanations regarding each pair  $(s,a)\in Q\times V$, the complexity of checking the condition of Claim 1 in the proof of Theorem 
\ref{decidable1} follows.

If all $[q,a]$ sets are empty, we go on to check the conditions of Claim 2 from the previous proof, i.e., for each $a\in V$, whether there exists an $a$-deficient cycle in a state $q$ reachable from $q_0$ and from which a final state can be reached while first reading some word $y$ visiting only $a$-deficient states and then reading at least one $a$. To check this, for each $a$, we first restrict the graph to $a$-deficient vertices and construct its transitive closure. Then, for each state $q$ having a cycle in the restriction, we check whether there is a path from $q$ to an accepting state, such that the path first visits only $a$-deficient states and then reads an $a$ from a state $r$. The number of possible candidates for $q$ is $\#(Q)$, the same holds for $r$, and for each candidate pair $q,r$ we need to check whether $q$ is reachable from $q_0$ (using the transitive closure), whether $r$ is reachable from $q$ visiting only $a$-deficient states (using the transitive closure of the restriction), and whether some final state $f$ is reachable from $r$ (one can remove states from which no final state is reachable after the construction of the transitive closure, making this step $O(1)$ at this stage). Altogether, checking the conditions of Claim 2 takes $\bigo(\#(V)\#(Q)^3)$ time, dominated by constructing the transitive closures for the restrictions.

\ep

Next we show that the previous decision algorithm is, in fact, enough to compute the asymptotic jump complexity of a given DFAwtl, due to the collapsing hierarchy.


\begin{theorem}\label{sublinear} 
	Let $f:\N\ra \N$ be a non-constant-bounded function such that 
	$\displaystyle{\lim_{n\rightarrow\infty}\frac{f(n)}{n}=0}$. There is no DFAwtl $M$ such that $JC_M(n)=f(n)$.
\end{theorem}

\bp
Assume the contrary, namely let $M=(Q,V,\delta,q_0,F)$ be a DFAwtl such that $JC_M(n)\notin \bigo(1)$ and $\lim_{n\rightarrow \infty}\frac{JC_M(n)}{n}=0$.
We consider a sufficiently large $n$ and let $w$ be a word in $L(M)$ of length $n$
such that $jc_M(w)=JC_M(n)$. For simplicity, we denote $JC_M(n)=m$. The word $w$ can be decomposed into $2m+1$ subwords, not necessarily nonempty, $x_1,x_2,\dots,x_{2m+1}$, such that $w$ can be recomposed by
$$w=x_1\diamond x_2\diamond x_3\dots x_{2m}\diamond x_{2m+1},$$
where each diamond represents either a position in $w$ when a jumping step is to be made or a letter which is consumed in a jumping step, therefore each diamond has to be replaced by either $\eps$ or a letter, satisfying the 
following conditions:

(i) $m$ diamonds are to be replaced by $\eps$ and the other $m$ by letters;

(ii) for any prefix, the number of diamonds replaced by letters is at most the number of diamonds replaced by $\eps$.  

Assume that the number of states of $Q$ is $k$. We construct another word $w'$, derived from $w$, as follows:

(i) Each segment $x_i$ is left unchanged, provided its length is smaller than $k$,
or it is replaced by a word $x'_i$, of length at most $k-1$ with the property that
if $q\in Q$ is the current state when $x_i$ is to be read, $\delta(q,x_i)=\delta(q,x'_i)$, without any jump. Clearly, this can be done by removing all the cycles of the path $\delta(q,x_i)$.

(ii) The letters inserted for reconstructing $w$ as above are inserted at the same positions for the construction of $w'$.

It follows that $w'\in L(M)$ and $|w'|\le (2m+1)(k-1)+m=m(2k-1)+k-1$.
Obviously, $jc_M(w')\le JC_M(m(2k-1)+k-1)$ must hold. On the other hand, 
from our construction $jc_M(w')=m=JC_M(n).$ However, we can argue like that about any word $w$ with $jc_M(w)=JC_M(n)=m$, which means that for any $n$ we have words $w'$ with length $\bigo(m)$ accepted with $m$ jumps, contradicting $\lim_{n\rightarrow \infty}\frac{JC_M(n)}{n}=0$.
\ep

This is in stark contrast with the nondeterministic case:

\begin{theorem}[\cite{MitranaPPS24}]
	There exists an NFAwtl $M$ such that $L(M)$ is a non-regular language, and $JC_M(n)\in \mathcal{O}(\sqrt n)$.
\end{theorem}

\begin{corollary}[\cite{MitranaPPS24}]
	$JCL(\log n)\setminus JCL(1)\ne\emptyset$.
\end{corollary}

Now let us prove an upper bound on the number of jumps made by DFAwtl with $\mathcal{O}(1)$ complexity, that will allow us to deduce an algorithm for deciding the language equivalence of those machines.

\begin{lemma}\label{lem:jumpupperbound}
    If DFAwtl $M=(Q,\Sigma, q_0,\delta, F)$ makes at least $\#(Q)$ jumps on some input, then $JC(M)\in \Omega(n)$.
\end{lemma}
\bp
    Suppose that there exists a word $w$ that is accepted by $M$ with at least $\#(Q)$ jumps, and consider the computation leading to its acceptance. The computation can be described by a sequence of position triples $(s_1, e_1, p_1), (s_2, e_2, p_2), \dots$, where each triple $(s_i, e_i, p_i)$
    means the machine reads $w[s_i .. e_i]$ sequentially after which it jumps and reads $w[p_i]$, the letter at position $p_i$. For simplicity we assume that the positions refer to the original input $w$, not to the remaining input.

    Since the machine makes at least $\#(Q)$ jumps, we get that there exist $i<j$ such that reading $w[s_1.. e_1]w[p_1]w[s_2.. e_2]w[p_2]\cdots w[s_i.. e_i]$ from the initial state leads to the same state $r$ as reading $w[s_1.. e_1]w[p_1]\cdots w[s_j..e_j]$. Let $i$ and $j$ be the smallest such values.
    
    We separate out the 
    jumping positions that are between the first and second time we reach the repeated state $r$, but are read before the state is reached for the first time:
    $$S_r = \{p_1,\dots, p_{i-1}\} \cap \{p \mid e_i < p < e_j\}.$$
    For instance, if the input $abcde$ is read in the order $adbce$ such that $$(q_0,abcde)\rightarrow (q_1,bcde)\rightarrow (q_2,bce)\rightarrow (r,ce)\rightarrow (q_3,e)\rightarrow (r,\varepsilon),$$
    then $d$ is between the letters read the first and second time $r$ is reached, but is actually consumed before $r$ is reached for the first time, so $S_r=\{4\}$.
    
    Furthermore, for a factor $w[i.. j]$ of $w$ and a set of positions $S$ in $w$, let us denote by $w[i.. j]_{-S}$ the word we obtain from $w[i.. j]$ by deleting the positions $S$:
    $$w[i.. j]_{-S} = \prod_{k\in [i.. j]\setminus S}w[k].$$

    Now we are ready to construct a sequence of words $v_n$ of length $O(n)$ that are accepted by $M$ with $\Omega(n)$ jumps. If $\prod_{k=i_1}^{j}w[s_k..e_k]\neq \varepsilon$, then let $x$ be the word we get if we insert $w[p_i]$ into the second position of $\prod_{k=i+1}^j(w[s_k..e_k]_{-S_r}\cdot w[p_k])$. We define
    $$v_n = w[1..e_j] x^n w[e_j+1..|w|].$$
    If $\prod_{k=i_1}^{j}w[s_k..e_k]=\varepsilon$, then $w[s_{j+1}]$ is translucent for each state on the path from $r$ when reading $\prod_{k=i+1}^j w[p_k]$, because those letters are read in jumps. In this case, let $x=\prod_{k=i+1}^j w[p_k]$ and we define 
    $$v_n = w[1..s_{j+1}] x^n w[s_{j+1}+1..|w|].$$
    The length of the word is $|w|+n|x|$ and in the accepting computation for $v_n$, the machine makes a jump each time it reads $x$, giving $n$ jumps.
\ep

Using the upper bound from the previous theorem, and the jump complexity reducing simulation for NFAwtl from~\cite{MitranaPPS24}, we can design an algorithm for testing the language equivalence of DFAwtl with jump complexity $\bigo(1)$.

\begin{theorem}
    The equivalence problem for DFAwtl of constant jump complexity is decidable.
\end{theorem}
\bp
First we recall that for each NFAwtl $A$ such that $JC_A(n)=k$, it is possible to construct NFAwtl $A'$ with $JC_{A'}(n)=k-1$, such that $L(A)=L(A')$~\cite{MitranaPPS24}. As DFAwtl are implicitly also NFAwtl, we can construct NFAwtl for them, too, decreasing the jump complexity by one. Now suppose that for two given DFAwtl $A$ and $B$ having $m$ and $n$ states, respectively, we want to decide whether $L(A)=L(B)$. We iterate the above mentioned NFAwtl construction $m$ and $n$ times, respectively, to obtain NFAwtl $A'$ and $B'$ such that $L(A)=L(A')$ and $L(B)=L(B')$. By Lemma~\ref{lem:jumpupperbound} this is enough to obtain NFAwtl with jump complexity $0$ accepting the same language as the input machines. Note that the resulting NFAwtl accept each word in their respective languages on at least one path without any jumps. Now consider the machines $A'$ and $B'$ as classical NFA: each accepting path without jumps is still an accepting path when treated as classical NFA, and each word in the language having one, we get that $L(A')\subseteq L_{NFA}(A')$, where $L_{NFA}(A')$ is the language accepted by $A'$ when treated as an NFA.
It is also immediate that any accepting path in the NFA $A'$ is an accepting path in the NFAwtl $A'$, so $L(A')=L_{NFA}(A')$. Therefore, after constructing the NFAs $A'$ and $B'$, we can check $L(A')=L(B')$ by classical methods and answer the decision problem accordingly.
\ep

Note that, due to the need of determinizing the resulting NFA before we can check equivalence, the resulting algorithm has exponential time complexity in the size of the input DFAwtl. We have no hardness results to present at the moment, and we propose to further investigate the open problem whether the equivalence of constant jump complexity DFAwtl is decidable in polynomial time.

\section{Deciding regularity for DFAwtl over binary alphabets}

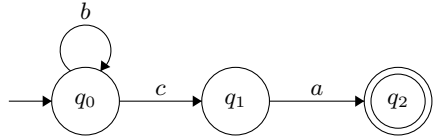
\begin{figure}
    \centering
    \begin{tikzpicture}[scale=0.15]
\tikzstyle{every node}+=[inner sep=0pt]
\draw [black] (13.3,-27.9) circle (3);
\draw (13.3,-27.9) node {$q_0$};
\draw [black] (26.6,-27.9) circle (3);
\draw (26.6,-27.9) node {$q_1$};
\draw [black] (41,-27.9) circle (3);
\draw (41,-27.9) node {$q_2$};
\draw [black] (41,-27.9) circle (2.4);
\draw [black] (11.977,-25.22) arc (234:-54:2.25);
\draw (13.3,-20.65) node [above] {$b$};
\fill [black] (14.62,-25.22) -- (15.5,-24.87) -- (14.69,-24.28);
\draw [black] (6.5,-27.9) -- (10.3,-27.9);
\fill [black] (10.3,-27.9) -- (9.5,-27.4) -- (9.5,-28.4);
\draw [black] (16.3,-27.9) -- (23.6,-27.9);
\fill [black] (23.6,-27.9) -- (22.8,-27.4) -- (22.8,-28.4);
\draw (19.95,-27.4) node [above] {$c$};
\draw [black] (29.6,-27.9) -- (38,-27.9);
\fill [black] (38,-27.9) -- (37.2,-27.4) -- (37.2,-28.4);
\draw (33.8,-27.4) node [above] {$a$};
\end{tikzpicture}
    \caption{DFAwtl accepting the regular language $ab^*c+b^*ac+b^*ca$ with jump complexity $\Theta(n)$, witnessed by inputs $ab^*c$.}
    \label{fig:regDFAwtl}
\end{figure}

In this section we look at deciding regularity of the language of a given DFAwtl. We do not have a characterization of regular languages accepted by DFAwtl in terms of jumping complexity, since there are DFAwtl accepting regular languages with linear jumping complexity (asymptotically maximal), as illustrated in Figure~\ref{fig:regDFAwtl}. 

However, not unexpectedly, analyzing the jumping behavior of DFAwtl is crucial in determining regularity. In what follows, we present a solution for DFAwtl over binary alphabets. The method may not generalize to larger alphabets, although we conjecture that in those cases regularity is still decidable, and the characterization on which a decision algorithm could be based, is similar in spirit to the binary case.

\begin{problem}\label{prob:binregularity}
    Given a DFAwtl $A$ with a binary input alphabet, is $L(M)$ regular?
\end{problem}


\begin{definition}
A circuit $p_0,\dots, p_n=p_0\in Q$ is \textbf{$a$-jumping} for some $a\in \Sigma$, if for each $i\in [n-1]$ such that $\delta(p_i,a)=p_{i+1}$, the state $p_i$ is $b$-deficient for some $b\in\Sigma$, and there is at least one $i$ such that $\delta(p_i,b)=p_{i+1}$ for $b\neq a$.    
\end{definition}
 As we will argue in the decision algorithm later, it is enough to consider circuits that are cycles, too, that is, do not visit states repeatedly, but for now we do not require it.

\begin{lemma}\label{thm:charbinreg}
    A DFAwtl $M=(Q,\{a,b\},q_0,\delta,F)$ accepts a non-regular language if and only if it has a reachable $x$-jumping circuit for some $x\in\{a,b\}$.
\end{lemma}
\bp
    For the if direction, w.l.o.g. suppose that a reachable $a$-jumping circuit $p_0,\dots, p_n$ in the automaton, such that $\delta(p_i,a_i)=p_{i+1}$ for letters $a_0,\dots,a_{n-1}\in\{a,b\}$. As the circuit can be shifted circularly, w.l.o.g. we may assume that $a_0=a$. This means that we can write $a_0\cdots a_{n-1}=ab^{k_1}ab^{k_2}\cdots ab^{k_\ell}$, for some $k_1,\dots,k_\ell$ such that $\ell+k_1+\cdots+k_\ell=n$. The circuit being $a$-jumping, we know that all states reading $a$ in this circuit, i.e., $p_0,p_{k_1+1},\dots$ are $b$-deficient. This means that $p_0 b^{n-\ell}a^\ell\vdash^* p_0$, because the $a$'s can be read with jumps. Further, we get $p_0 (b^{n-\ell})^i (a^\ell)^i\vdash^* p_0$ for all $i\geq 0$. Now consider the language $L'=L(M)\cap u(b^{n-\ell})^+a^+v$, where $u,v$ are some words such that $\delta(q_0,u)=p_0$, $\delta(p_0,v)\in F$ without jumps. Suppose that $L(M)$ is regular, implying that $L'$ must be as well, and by the previous argument we know that $u(b^{n-\ell})^ia^iv\in L'$, for any $i\geq 0$. Taking an $i$ large enough that the length of $(b^{n-\ell})^i$ is larger than the number of states accepting $L'$, we get that there is some $r>0$ such that $u(b^{n-\ell})^i(b^r)^+a^iv\subset L'$. Now taking the word $u(b^{n-\ell})^i(b^r)^{(n-\ell)|v|}a^iv$ which should be in $L'$, we notice that after reading $u$ and looping in the circuit for $i$ times, then looping some more until all the $a$'s from $v$ are also read from the tape, $M$ ends up in one of the $b$-deficient states $p_0,p_{k_1+1}, \dots$ with only $b$'s left on the tape, rejecting the input and contradicting the assumption that $L'$ is regular. From here, $L(M)$ cannot be regular.

    For the only if direction, we assume that $M$ has no $x$-jumping circuit for any $x\in \{a,b\}$, and we will construct an $\varepsilon$-NFA $M'$ that simulates $M$, defined as follows:
    \begin{itemize}
        \item $M'=(Q', \{a,b\}, \delta', (q_0,0,0), F')$.
        \item $Q'=\{(q,m,n) \mid q\in Q \text{ and } m, n\in \{0,\dots, |Q|-1\}\}$; the second and third component of states will remember the number of jumps performed while reading $a$'s and $b$'s, respectively.
        \item $F'=\{(f,0,0) \mid f\in F\}$;
        \item the transitions of $M'$ will be of three types, recording jumps (1), amortizing jumps (2), sequential transitions inherited from $M$ (3). In detail:
        \begin{enumerate}
            \item for each $a$-deficient state $q$, such that $\delta(q,b)=p$ we define $\delta'((q,m,n),\varepsilon)=(p,m,n+1)$; we define the transitions jumping over $b$'s analogously.
            \item for each state $q$ and $n\geq 1$, we define $\delta'((q,m,n),b)=(q,m,n-1)$; for each state $q$ and $m\geq 1$, we define $\delta'((q,m,n),a)=(q,m-1,n)$;
            \item for each state $q$, and each $n$, we define $\delta'((q,0,n),a)=(p,0,n)$ if $\delta(q,a)=p$, and $\delta'((q,n,0),b)=(p,n,0)$ if $\delta(q,b)=p$.
        \end{enumerate}
    \end{itemize}

The constructed $\varepsilon$-NFA simulates $M$'s computations. Whenever a jump occurs reading $x$ during a computation of $M$, the NFA $M'$ will perform the state change of $M$ recording it in the first component and increment the respective counter. The next time it encounters $x$ on the tape, it will decrement it without changing the first component tracking the state of $M$. The final states have counters set to $0$, meaning all jumps have been accounted for. The key for the finite state simulation is that the automaton will never need to remember more than $\#(Q)-1$ jumps that have not been accounted for yet. This is ensured by the fact that in any circuit of $M$ having $y$-deficient states, since the circuit is not $x$-jumping, after at most $\#(Q)-1$ transitions we encounter an $x$-transition from a state $p$ for which $\delta(p,y)$ is also defined. Continuing with that $x$-transition means reading an $x$ sequentially, i.e., $x$ must be the leftmost symbol left on the tape. This implies that any $x$ that has been read earlier while jumping, must occur before this position. The simulating machine, therefore, can keep recording jumps on reading $x$ and accounting for all of them no later than $\#(Q)-1$ steps after the first one was recorded.
\ep

\begin{theorem}\label{thm:decregdec}
    Problem~\ref{prob:binregularity} is decidable in $\bigo(n^4)$ time, where $n$ is the number of states in the input DFAwtl.
\end{theorem}
\bp
    Let the input DFAwtl be $A=(Q, \{a,b\}, \delta, q_0, F)$ with $\#(Q)=n$. Removing unreachable states and states from which no final state is reachable (except, possibly, a sink state) can be done in time $\bigo(n^3)$ by, e.g., the Floyd-Warshall algorithm~\cite{Floyd}, so from here on we may assume that $A$ has only useful states.

    From the characterization of DFAwtl accepting binary regular languages in Lemma~\ref{thm:charbinreg}, we get that it is enough to check whether $x$-jumping circuits exist in $A$. First we note that, in fact, it is enough to check whether $x$-jumping cycles exist. Suppose that there is some $x$-jumping circuit $p_0,\dots, p_n\in Q$ for $x\in \{a,b\}$, so for each $i$ such that $\delta(p_i,x)=p_{i+1}$, the state $p_i$ is $y$-deficient for $y\neq x$, and there is at least one $i$ such that $\delta(p_i,y)=p_{i+1}$. Let us also assume that there are $i,j\in [n]$ such that $i\neq j$ but $p_i=p_j$. We get two circuits $C_1=p_0,\dots,p_i,p_{j+1},\dots,p_n$ and $C_2=p_{i},\dots,p_{j}$. If $C_1=C_2$, then removing $C_2$ still leaves an $x$-jumping circuit, and we repeat the argument with the reduced circuit. Otherwise, $C_1$ and $C_2$ share some states (at least $p_i$), but not all states. As $A$ is deterministic, there is a shared state $p_k$ for some $k\in [n]$, from which $x$ follows $C_1$ and $y$ follows $C_2$ (or the other way around). This means that the original circuit cannot be $x$-jumping, because $y$ is not translucent for $p_k$ and $\delta(p_k,x)=p_{k+1}$. From this contradiction we conclude that if $A$ has an $x$-jumping circuit, it must have an $x$-jumping cycle.

    Detecting $x$-jumping cycles can be done by dynamic programming. First, we note that since $x$-transitions from not $y$-deficient states $p$ are never parts of such cycles, we may remove those $\delta(p,x)$ transitions from the transition graph. Let the transition function having only the remaining transitions be $\delta'$. From there, the question reduces to checking whether there exists a cycle having at least one transition for each of the letters. We construct a $4$-dimensional matrix $M$ indexed by step count, pairs of states and the alphabet. For any $i\in [n]$, $p,q\in Q$ and $a\in \Sigma$ we want $M[i,p,q,a]=1$ if and only if there exists a path of length at most $i$ from state $p$ to $q$ such that there is at least one transition on the path labeled by $a$. We initialize the whole matrix with $0$'s. In the inductive step, suppose that the matrix is computed up to step $i$, that is, all entries $M[j,p,q,a]$ with $j\leq i$. There are three possibilities to set $M[i+1,p,q,a]$ to $1$:
    \begin{enumerate}
        \item[(1)] if $M[i,p,q,a]=1$, or
        \item[(2)] if $\delta'(p,a)=q$, or
        \item[(3)] if there exists some $r\in Q$ such that $M[i,p,r,b]=1$  and $\delta'(r,a)=q$.
    \end{enumerate}
    We stop computing $M$ when $i>\#(Q)$. From the construction it is clear that indeed $M[i,p,q,a]=1$ if and only from $p$ to $q$ there is some path of length at most $i$, having at least one $a$ transition. The last step is to check whether there exist some $p,q,r\in Q$ such that $M[n,p,r,a]=1$ and $M[n,r,q,b]=1$.

    The size of the matrix is $\bigo(n^3)$ (since the alphabet is fixed with $\#(V)=2$), and to compute each entry we need to check conditions (1)-(3) from above. The first two can be checked in $\bigo(1)$ time and (3) can be checked in $\bigo(n)$ time, iterating over $r$. Constructing $M$, hence, takes $\bigo(n^4)$ time. To perform the last step, we need to iterate over triples $p,q,r\in Q$ and check in each iteration the corresponding matrix entry (in $\bigo(1)$ time), which yields complexity $\bigo(n^3)$ for the last step. Altogether, the time complexity of the algorithm is dominated by constructing $M$, so it is $\bigo(n^4)$.    
\ep

\section{Conclusion}
The jump complexity landscape of DFAwtl is in stark contrast both with NFAwtl and with  deterministic one-way jumping automata. As we have proved here, DFAwtl only has two separable asymptotic jump complexity classes, $\bigo(1)$ and $\bigo(n)$, whereas for NFAwtl we also have separable classes $\bigo(\log n)$ and $\bigo(\sqrt{n})$, and deterministic one-way automata, too, have a separable $\bigo(\log n)$ sweep complexity class (a measure similar to jump complexity). Due to the collapsing jump complexity hierarchy for DFAwtl, we were able to obtain a polynomial time algorithm for computing the asymptotic jump complexity of a given input machine, a problem which is still open for the other two aforementioned models. Upper bounding the number of jumps possible in DFAwtl with $\bigo(1)$ jump complexity made it possible to decide equivalence for two such machines. This suggests that it is possible that the minimal DFAwtl is computable for constant complexity inputs. We think that is also a very interesting problem, which could lead to other directions of inquiry regarding, for instance, descriptional complexity. A related research objective worth considering is to find a better way to decide equivalence, as the current algorithm runs in exponential time in the worst case due to the built-in NFA determinization step.

Finally, deciding regularity is one of the most fundamental problem considered for all such non-sequentially processing automata models. The problem is undecidable for NFAwtl~\cite{translucent} and nothing is known about the case of one-way jumping automata. Here we showed that at least over binary alphabets, the problem is decidable for DFAwtl. We conjecture that this is still the case for larger alphabets, perhaps provable by adapting the characterization obtained in Lemma~\ref{thm:charbinreg}. 

\bibliographystyle{splncs04}
\bibliography{jumpcomplexity}

\end{document}